\documentclass[11pt]{article}
\pdfoutput=1
\usepackage{amsmath,amsfonts,amssymb,color}
\usepackage{cite,url}
\usepackage{graphicx}
\usepackage[utf8]{inputenc} 
\usepackage[bottom]{footmisc}
\usepackage{hyperref}

\allowdisplaybreaks

\def\msk2lam{$m_s\approx 5.5$~TeV, $m_a\approx 4.2$~TeV}

\def\smallSM{{\rm{\scriptscriptstyle SM}}}
\def\smallBSM{{\rm{\scriptscriptstyle BSM}}}

\def\smallGW{{\rm{\scriptscriptstyle GW}}}

\def\beq{\begin{equation}}
\def\eeq{\end{equation}}
\def\bea{\begin{eqnarray}}
\def\eea{\end{eqnarray}}
\def\nn{\nonumber}
\def\wt{\widetilde}

\def\Theavy{\widetilde T}

\def\mZq{m^2_Z}
\def\mWq{m^2_W}
\def\mHq{m^2_H}
\def\mAq{m^2_A}
\def\mHpq{m^2_{H^\pm}}

\def\sdbe{s_{2\beta}}
\def\cdbe{c_{2\beta}}
\def\sbe{s_{\beta}}
\def\cbe{c_{\beta}}

\def\sq2{\sqrt{2}}

\def\msbar{\overline{\rm MS}}

\long\def\symbolfootnote[#1]#2{\begingroup%
\def\thefootnote{\fnsymbol{footnote}}\footnote[#1]{#2}\endgroup}

\makeatletter
\newcommand{\vast}{\bBigg@{3}}
\makeatother

\begin{document}

\begin{titlepage}

\begin{center}

\vspace{1cm}

{\LARGE \bf Revisiting the Higgs-mass calculation }\\[3mm]

{\LARGE \bf in the scale-invariant THDM} 

\vspace{1cm}

{\Large Pietro~Slavich$^{\,a}$}

\vspace*{1cm}

{\sl ${}^a$
   Sorbonne Université, CNRS,
  Laboratoire de Physique Th\'eorique et Hautes Energies, 
 
  4 Place Jussieu, F-75005, Paris, France.}
\end{center}
\symbolfootnote[0]{{\tt e-mail:}}
\symbolfootnote[0]{{\tt slavich@lpthe.jussieu.fr}}

\vspace{0.7cm}

\abstract{We revisit the one-loop calculation of the Higgs-mass
  spectrum of the scale-invariant THDM, relying on a direct
  calculation of the relevant Feynman diagrams. We highlight a number
  of incorrect assumptions in earlier calculations that relied on the
  effective-potential approach. In contrast with the earlier findings,
  we show that the one-loop corrections can have an effect of ${\cal
    O}(10\%)$ on the predictions for the BSM-Higgs masses, and they
  can also induce non-negligible mixing between the SM-like and BSM
  states in the neutral-scalar sector.}

\vfill

\end{titlepage}


\setcounter{footnote}{0}

\section{Introduction}
\label{sec:intro}
The discovery of a Higgs boson with mass around
$125$~GeV~\cite{CMS:2012qbp, ATLAS:2012yve} and properties compatible
with the predictions of the Standard Model
(SM)~\cite{ParticleDataGroup:2024cfk} goes a long way toward
elucidating the mechanism of electroweak (EW) symmetry breaking, but
does not by itself preclude the existence of additional, beyond-the-SM
(BSM) Higgs bosons with masses around or even below the TeV scale,
which could still be discovered in the current or future runs of the
Large Hadron Collider (LHC).

The Two-Higgs-Doublet Model (THDM) is one of the simplest and
best-studied extensions of the SM (for reviews see, e.g.,
refs.~\cite{Gunion:1989we,Aoki:2009ha,Branco:2011iw}). In the
CP-conserving versions of the model, the Higgs sector includes five
physical states: two CP-even scalars, $h$ and $H$; one CP-odd scalar,
$A$; and two charged scalars, $H^\pm$. In general realizations of the
THDM, all of the scalar masses can be treated as free parameters, and
the couplings of both $h$ and $H$ to fermions and gauge bosons deviate
from the SM predictions. As discussed, e.g., in
ref.~\cite{Gunion:2002zf}, the so-called ``alignment'' condition -- in
which one of the CP-even scalars has SM-like couplings to fermions and
gauge bosons -- can be realized through decoupling, when all of the
other Higgs bosons are much heavier, or without decoupling, when a
specific configuration of parameters in the Lagrangian suppresses the
mixing between the SM-like scalar and the other CP-even scalar.

A version of the THDM that features (approximate) alignment without
decoupling is the so-called ``Scale-Invariant THDM'' (SI-THDM)
introduced in ref.~\cite{Lee:2012jn}. In this model there are no
explicit mass parameters in the scalar potential, rendering the
classical action of the theory scale invariant. However, quantum
corrections to the scalar potential break both the scale invariance
and the EW symmetry, inducing a vacuum expectation value (vev) $v$ for
the SM-like Higgs field, and thus masses proportional to combinations
of couplings times $v$ for all of the particles in the model. This
mechanism of radiative symmetry breaking was first discussed by
Coleman and E.~Weinberg (CW) in ref.~\cite{Coleman:1973jx}, and later
extended to models with additional scalar fields by Gildener and
S.~Weinberg (GW) in ref.~\cite{Gildener:1976ih}. Owing to the reduced
number of free parameters w.r.t.~the usual THDM -- namely, the three
mass parameters for the Higgs doublets are set to zero -- the Higgs
sector of the SI-THDM is particularly constrained. Even after $v$ is
generated radiatively, one of the CP-even scalars, which we identify
with $h$, remains massless at the tree level,\footnote{With a common
abuse of language, we refer to a ``tree-level'' mass spectrum even if
$v$ itself is induced at one loop.} has SM-like couplings to fermions
and gauge bosons, and does not mix with the other CP-even scalar. A
mass for this SM-like Higgs boson, $M_h$, is generated at one loop,
and a sum rule connects it with all of the other particle
masses. Thus, if $M_h\approx 125$~GeV is treated as an input, the sum
rule constrains one of the masses of the BSM Higgs bosons.

In view of the crucial role played by radiative corrections in the
SI-THDM, several studies have been devoted over the years to the
precise determination of its Higgs-mass spectrum. After the original
one-loop calculation of ref.~\cite{Lee:2012jn}, the one-loop
corrections to the mass matrix for the CP-even scalars were revisited
by different groups in refs.~\cite{Lane:2018ycs, Lane:2019dbc,
  Eichten:2021qbm}. Furthermore, the dominant two-loop corrections
were computed: first along the $h$ direction in
ref.~\cite{Braathen:2020vwo}, and later for the full mass matrix in
ref.~\cite{Eichten:2022vys}.  All of these calculations relied on the
effective-potential approach, which neglects the external-momentum
effects, and the two-loop calculations also neglected the effects
controlled by the EW gauge couplings. However, the one-loop
calculations presented in refs.~\cite{Lee:2012jn, Lane:2018ycs,
  Lane:2019dbc, Eichten:2021qbm} appear to be incorrect. First of all,
they find that there are no one-loop corrections to the masses of $A$
and $H^\pm$. As a result, the predictions for those masses, which do
not vanish at the tree level, would be both scale and gauge dependent
at the one-loop order. Furthermore, those calculations find that the
only one-loop contributions to the mixing between $h$ and $H$ arise
from diagrams involving the top quark, despite the presence in the
Lagrangian of both $h\Phi\Phi$ and $H\Phi\Phi$ couplings (where
$\Phi=H,A,H^\pm$ denotes the BSM Higgs bosons). A closer inspection of
the effective-potential calculations of refs.~\cite{Lee:2012jn,
  Lane:2018ycs, Lane:2019dbc, Eichten:2021qbm} reveals that both of
these problems stem from an incorrect assumption for the
field-dependent masses of BSM-Higgs and gauge bosons, which does not
properly account for their dependence on the fields other than $h$. We
remark that this problem also affects the two-loop calculation of
ref.~\cite{Eichten:2022vys}, but it does not affect the two-loop
calculation of ref.~\cite{Braathen:2020vwo}, which was performed only
along the $h$ direction.\footnote{For completeness, we mention here
two additional recent studies of the SI-THDM, refs.~\cite{Nhi:2025iob,
  Baouche:2025jsf}, which also feature one-loop calculations of the
Higgs masses in the effective-potential approach. As will be discussed
later, the treatment of the minimum conditions for the scalar
potential in ref.~\cite{Nhi:2025iob} imposes an arbitrary restriction
on the parameter space of the model.  Ref.~\cite{Baouche:2025jsf} has
even broader issues, stemming in part from an incorrect identification
of the SM-like state at the tree level. Neither of these studies
specifies the definitions used for the field-dependent masses.}

Due to the presence in the Higgs sector of the SI-THDM of four neutral
and four charged scalar fields that interact with each other --
counting also the would-be-Goldstone bosons, $G^0$ and $G^\pm$ -- the
correct determination of the field-dependent masses required in the
effective-potential calculation is not trivial (see
ref.~\cite{Pilaftsis:2024uub} for a formal approach to this issue). In
this paper, we bypass all complications related to the field-dependent
masses by computing directly all of the Feynman diagrams that are
relevant to the one-loop calculation of the Higgs-mass spectrum of the
SI-THDM. Besides correcting the errors in the literature, this
approach allows us to account for the external-momentum effects in the
self-energies of the BSM Higgs bosons, eventually resulting in
predictions for the masses that are both scale and gauge independent
up to the perturbative order considered in our calculation.

The rest of the article is organized as follows: in
section~\ref{sec:HiggsSector} we first introduce the Higgs sector of
the SI-THDM at the tree level, and then describe our calculation of
the one-loop corrections to the Higgs masses; in
section~\ref{sec:numerical} we illustrate the numerical impact of the
corrections to the Higgs masses in two representative scenarios;
section~\ref{sec:conclusions} contains our conclusions. Finally, the
appendix~A contains explicit formulas for all of the one-loop
quantities relevant to the Higgs-mass calculation, and in the
appendix~B we discuss the role of the so-called Gildener-Weinberg
scale introduced in ref.~\cite{Gildener:1976ih}.
 
\section{The Higgs sector of the SI-THDM}
\label{sec:HiggsSector}

We start this section by describing the ``tree-level'' mass spectrum
of the SI-THDM in the presence of a Higgs vev $v$ generated
radiatively through the CW/GW mechanism. We then describe the full one-loop
calculation of the masses of all of the physical scalars in the model. 

\newpage

\subsection{The scalar potential and the Higgs-mass spectrum at the tree level}
\label{sec:tree}

We consider a version of the THDM where flavor-changing
neutral-current interactions are forbidden at the tree level by a
$Z_2$ symmetry. In the basis where this $Z_2$ symmetry applies, the
scalar potential can be parametrized as
\bea
V_0 &=& \frac{\lambda_1}{2}\left(\Phi_1^\dagger \Phi_1\right)^2
+~ \frac{\lambda_2}{2}\left(\Phi_2^\dagger \Phi_2\right)^2
+~ \lambda_3\,\Phi_1^\dagger \Phi_1\,\Phi_2^\dagger \Phi_2
~+~ \lambda_4 \,\Phi_1^\dagger \Phi_2\,\Phi_2^\dagger \Phi_1
~+~ \frac{\lambda_5}{2}\left[\left(\Phi_1^\dagger \Phi_2\right)^2 +~ {\rm h.c.}\right]~,\nn\\
\label{eq:Vstd}
\eea
where all of the quartic couplings are assumed to be real to ensure CP
conservation. We do not need here to distinguish between different
THDM ``types'' according to the form of their Higgs--fermion
interactions, because in our calculation of the one-loop corrections
to the Higgs masses we neglect all Yukawa couplings except the one of
the top quark, which we assume to couple only to $\Phi_2$. We
decompose the two $SU(2)$ doublets as
\beq
\label{eq:Phik}
\Phi_k ~=~ \frac{1}{\sqrt2} \left(\!\begin{array}{c} \sqrt2\,\phi^+_k \\ 
v_k + \phi^0_k + i\,a_k  \end{array}\!\right)~~~~~(k=1,2)~,
\eeq
where the two (real) vevs are related by $v_1^2+v_2^2=v^2$, with
$v\approx 246$~GeV, and we define $\tan\beta \equiv v_2/v_1$. In the
absence of explicit mass parameters, the minimum conditions for the
scalar potential when $v\neq 0$ reduce to relations between the
quartic couplings and $\tan\beta$:
\beq
\label{eq:mintree}
\lambda_1 ~=~ - \lambda_{345}\,\tan^2\beta~,~~~~~~
\lambda_2 ~=~ - \lambda_{345}\,\cot^2\beta~,
\eeq
where we define $\lambda_{345}\equiv \lambda_3+\lambda_4+\lambda_5$
and, for later convenience, $\lambda_{45}\equiv
\lambda_4+\lambda_5$. In this constrained version of the THDM, the
mass matrices for the scalar, pseudoscalar and charged components of
the Higgs doublets in eq.~(\ref{eq:Phik}) are all diagonalized by the
same rotation $R(\beta)$, defined as
\beq
R(\beta) ~\equiv~\left(\!\begin{array}{rr} c_\beta&\!s_\beta\\-s_\beta
&\! c_\beta \end{array}\!\right)~,
\eeq
where we introduce the shortcuts $c_\theta \equiv \cos\theta$ and
$s_\theta\equiv \sin\theta$ for a generic angle $\theta$.  It is thus
convenient to rotate the whole Higgs doublets from the $Z_2$-symmetric
basis to the so-called Higgs basis:
\beq
\label{eq:hb}
\left(\!\begin{array}{c} \Phi_{\smallSM} \\ \Phi_{\smallBSM} \end{array}\!\right)
~=~ R(\beta)\,\left(\!\begin{array}{c}
\Phi_1 \\ \Phi_2 \end{array}\!\right)~,
\eeq
in which one of the doublets develops the full SM-like vev $v$ and the
other has vanishing vev:\footnote{Note that in this paper the
convention for $H$ differs from the one in
refs.~\cite{Degrassi:2023eii, Degrassi:2025pqt}: here we define
$H=\phi_\smallBSM^0$, whereas in our earlier papers on the aligned
THDM we had $H\rightarrow - \phi_\smallBSM^0$ in the alignment limit.}
\beq
\Phi_\smallSM ~=~ \left(\!\begin{array}{c} G^+ \\ \frac{1}{\sqrt2}
(v + h + i\,G^0)  \end{array}\!\right)~,~~~~~
\Phi_\smallBSM ~=~ \left(\!\begin{array}{c} H^+ \\ \frac{1}{\sqrt2}
(H + i\,A)  \end{array}\!\right)~.
\eeq
Using the minimum conditions in eq.~(\ref{eq:mintree}), we get the
tree-level masses:
\beq
\label{eq:massestree}
m^2_h \,=\, m^2_{G^0} \,=\, m^2_{G^\pm}=0~,~~~
m_H^2 = -\, \lambda_{345}\,v^2~,~~~
m_A^2 = -\, \lambda_5\,v^2~,~~~
m_{H^\pm}^2 = -\, \frac{\lambda_{45}}{2}\,v^2~.
\eeq
In practice, $h$ is the massless state -- called ``scalon'' in
ref.~\cite{Gildener:1976ih} -- associated with the flat direction in
field space along which the vev $v$ is generated by the quantum
corrections to the scalar potential. Being aligned in field space with
the vev, $h$ has SM-like couplings to gauge bosons and to fermions. In
contrast, the three BSM Higgs bosons acquire squared masses
proportional to $v^2$ times combinations of the three quartic
couplings $\lambda_3$, $\lambda_4$, and $\lambda_5$. Their trilinear
couplings to pairs of gauge bosons vanish, and their couplings to the
top quark are all proportional to $\cot\beta$.

While the tree-level alignment of the Higgs sector of the SI-THDM is a
welcome feature, the masslessness of the SM-like Higgs boson is
obviously unrealistic. Luckily, as we will see in the next section, a
suitable mass for $h$ can be generated by the one-loop corrections,
albeit at the price of some misalignment from the SM-like direction.

\subsection{One-loop corrections to the Higgs-boson masses}

Beyond the tree level, the minimum conditions for the effective
potential become
\beq
\label{eq:minloop}
\lambda_1 ~=~ - \lambda_{345}\,\tan^2\beta
\,-\,\frac{2}{v^3c_\beta^2}\,\left(T_h - T_H\,\tan\beta\right)~,~~~~~~
\lambda_2 ~=~ - \lambda_{345}\,\cot^2\beta
\,-\,\frac{2}{v^3s_\beta^2}\,\left(T_h + T_H\,\cot\beta\right)~,
\eeq
where $\tan\beta$ and the five couplings $\lambda_i$ are now
interpreted as $\msbar$-renormalized parameters at some scale $Q$.
The quantities $T_{\varphi}$ denote the finite parts of the one-loop
tadpole diagrams\,\footnote{Decomposing the one-loop effective
potential as $V_0 + \Delta V$, we also have $T_\varphi \,=\, d \Delta
V/d\varphi\,|_{\rm min}\,$.}  for the fields $\varphi = (h,H)$. By
making use of the minimum conditions in eq.~(\ref{eq:minloop}), we can
write the loop-corrected mass matrices for the scalar $(h,H)$,
pseudoscalar $(G^0,A)$, and charged $(G^\pm,H^\pm)$ sectors as
\bea
\label{eq:MS2p}
{\cal M}^2_S(p^2) &=&
\left(\!\begin{array}{cc}0&0\\0& -\, \lambda_{345}\,v^2\end{array}\right)
~+~
\left(\!\begin{array}{cc}
\Pi_{hh}(p^2) &
\Pi_{hH}(p^2)\\[2mm]
  \Pi_{hH}(p^2)&
 \Pi_{HH}(p^2) 
\end{array}\right)
\,-~ \frac{3}{v}\,
\left(\!\begin{array}{cc}
T_{h} & T_{H}\\[2mm]
T_{H} & \Theavy   
\end{array}\right)~,\\[3mm]
\label{eq:MP2p}
{\cal M}^2_P(p^2) &=&
\left(\!\begin{array}{cc}0&0\\0& -\, \lambda_5\,v^2\end{array}\right)
~+~
\left(\!\begin{array}{cc}
\Pi_{G^0G^0}(p^2) &
\Pi_{G^0A}(p^2)\\[2mm]
  \Pi_{G^0A}(p^2)&
 \Pi_{AA}(p^2) 
\end{array}\right)
\,-~ \frac{1}{v}\,
\left(\!\begin{array}{cc}
T_{h} & T_{H}\\[2mm]
T_{H} & \Theavy   
\end{array}\right)~,\\[3mm]
{\cal M}^2_C(p^2) &=&
\left(\!\begin{array}{cc}0&0\\
0& -\,\frac{\lambda_{45}}2\,v^2\end{array}\right)
~+~
\left(\!\begin{array}{cc}
\Pi_{G^{\pm}G^{\mp}}(p^2) &
\Pi_{G^{\pm}H^{\mp}}(p^2)\\[2mm]
  \Pi_{G^{\pm}H^{\mp}}(p^2)&
 \Pi_{H^{\pm}H^{\mp}}(p^2) 
\end{array}\right)
\,-~ \frac{1}{v}\,
\left(\!\begin{array}{cc}
T_{h} & T_{H}\\[2mm]
T_{H} & \Theavy   
\end{array}\right)~,
\label{eq:MC2p}
\eea
where $\Pi_{\varphi\varphi^\prime}(p^2)$ are the finite parts of the
one-loop self-energy matrices for the fields
$(\varphi,\varphi^\prime)$, $p^2$ is the external momentum, and we
defined $\Theavy = T_h + 2\cot2\beta\,T_H$. Again, the squared vev
$v^2$ and the couplings $\lambda_i$ entering the tree-level parts of
the mass matrices are now interpreted as $\msbar$-renormalized
parameters at some scale $Q$.  We remark that the identities
$\Pi_{G^0G^0}(0) = \Pi_{G^{\pm}G^{\mp}}(0) = T_h/v$ and $\Pi_{G^0A}(0)
= \Pi_{G^{\pm}H^{\mp}}(0) = T_H/v$ ensure that the corrections to the
masses and mixing of the would-be-Goldstone bosons vanish at
$p^2=0$. For what concerns the physical states, we note that the
mixing terms in the mass matrices of
eqs.~(\ref{eq:MS2p})--(\ref{eq:MC2p}) are purely of one-loop order,
and thus affect the mass eigenvalues only starting from the two-loop
order. Therefore, in a strict one-loop calculation, the ``pole'' mass
for a given physical state can be identified with the real part of the
corresponding diagonal element of the mass matrix, evaluated at an
external momentum equal to the tree-level mass of the particle
itself. For convenience, we also choose to replace the
$\msbar$-renormalized Higgs vev $v$, which is both scale and
gauge dependent, with the Fermi constant $G_F$, the two being related
at the tree level by $v = (\sqrt2G_F)^{-1/2}$.  We can thus
write:\footnote{Here and thereafter, we use $M_\varphi^2$ to denote
the squared pole mass of a scalar $\varphi$.  We instead use
$m_\varphi^2$ for the tree-level parts of
eqs.~(\ref{eq:MH1loop})--(\ref{eq:MHp1loop}), as well as when the
precise definition of the mass amounts to a higher-order effect.}
\bea
\label{eq:Mh1loop}
M_h^2 &=& \Pi_{hh}(0) \,-\, 3\,\frac{T_h}v~, \\[2mm]
\label{eq:MH1loop}
M_H^2 &=& -\frac{\lambda_{345}}{\sqrt2 G_F} \,-\, m_H^2\,\frac{\delta v^2}{v^2}
\,+\,{\rm Re}\,\Pi_{HH}(m_H^2) \,-\, 3\,\frac{\Theavy}v~,\\[2mm]
\label{eq:MA1loop}
M_A^2 &=& -\frac{\lambda_{5}}{\sqrt2 G_F} \,-\, m_A^2\,\frac{\delta v^2}{v^2}
\,+\,{\rm Re}\,\Pi_{AA}(m_A^2) \,-\, \frac{\Theavy}v~,\\[2mm]
\label{eq:MHp1loop}
M_{H^\pm}^2 &=& -\frac{\lambda_{45}}{2\sqrt2 G_F} \,-\, m_{H^\pm}^2\,\frac{\delta v^2}{v^2}
\,+\,{\rm Re}\,\Pi_{H^{\pm}H^{\mp}}(m_{H^\pm}^2) \,-\, \frac{\Theavy}v~,
\eea
where we defined $(\sqrt2 G_F)^{-1} = v^2 + \delta v^2$. We computed
all of the one-loop tadpoles and self-energies entering
eqs.~(\ref{eq:MS2p})--(\ref{eq:MHp1loop}) in a generic $R_\xi$ gauge
with the help of {\tt FeynArts}~\cite{Hahn:2000kx}. We also compared
our results for the one-loop self-energies with those obtained by
adapting the general results of ref.~\cite{Martin:2003it} to the
SI-THDM, and found agreement except for terms proportional to $\xi^2$
in $\Pi_{G^0G^0}(p^2)$ and $\Pi_{G^{\pm}G^{\mp}}(p^2)$ (those are
related to an issue in ref.~\cite{Martin:2003it}, already noticed in
ref.~\cite{Goodsell:2019zfs}, concerning the couplings of the
would-be-Goldstone bosons to ghost pairs). For $\delta v^2$ we
combined the SM contributions from refs.~\cite{Marciano:1980pb,
  Sirlin:1985ux, Degrassi:1992ff} with the BSM-Higgs contribution to
the self-energy of the $W$ boson.
As a nontrivial check of the correctness of our results,
we find that all of the pole-mass predictions in
eqs.~(\ref{eq:Mh1loop})--(\ref{eq:MHp1loop}) are both gauge and
scale independent at the one-loop order, once we take into account the
scale dependence of the $\msbar$-renormalized quartic couplings
entering the tree-level parts of the BSM-Higgs masses.
Explicit formulas for all of the one-loop quantities relevant to the
Higgs-mass calculation in the SI-THDM are provided in the appendix~A.

Combining the different terms in eqs.~(\ref{eq:MA1loop}) and
(\ref{eq:MHp1loop}) we find that the one-loop corrections to the
masses of $A$ and $H^\pm$ do not vanish, contrary to what was found in
refs.~\cite{Lee:2012jn, Lane:2018ycs, Lane:2019dbc}.  The finding in
refs.~\cite{Eichten:2021qbm, Eichten:2022vys} that one-loop
contributions to $M_H^2$ arise only from diagrams involving the top
quark is also incorrect.  All of this remains the case even if we
neglect the external-momentum effects and the connection of $v$ with
$G_F$, as was done in those papers. As mentioned in
section~\ref{sec:intro}, these discrepancies appear to stem from an
unjustified assumption in refs.~\cite{Lee:2012jn, Lane:2018ycs,
  Lane:2019dbc, Eichten:2021qbm, Eichten:2022vys} on the structure of
the field-dependent masses of the BSM-Higgs and gauge bosons, which
are all taken to be proportional to $(\Phi_1^\dagger\Phi_1 +
\Phi_2^\dagger\Phi_2)$ or, equivalently, to
$(\Phi_\smallSM^\dagger\Phi_\smallSM +
\Phi_\smallBSM^\dagger\Phi_\smallBSM)$. It is easy to see how, for
example, this assumption would incorrectly imply that the BSM Higgs
bosons do not contribute at all to $T_H$ and $\Pi_{hH}(0)$, because it
misses terms proportional to $hH$ in their field-dependent masses.

For the mass of the SM-like Higgs boson, see eq.~(\ref{eq:Mh1loop}),
our result agrees with the earlier papers:
\beq
\label{eq:Mhsum}
M_h^2 ~=~ \frac{1}{8\pi^2v^2}\, \left(m_H^4 + m_A^4 + 2\, m_{H^\pm}^4
+6\, m_W^4 + 3\,m_Z^4 - 4\,N_c\,m_t^4\right)~,
\eeq
where $N_c=3$ is a color factor. However, for the mixing term in the
mass matrix for the scalar sector at vanishing external momentum we find:
\beq
\label{eq:MS12}
\left[{\cal M}^2_S(0)\right]_{12} ~=~
\frac{1}{8\pi^2v^2}\, \left[m_H^2\,\left(3\,m_H^2+m_A^2+2\,m_{H^\pm}^2\right)
\,\cot2\beta -  4\,N_c\,m_t^4\,\cot\beta\right]~,
\eeq
to be contrasted with the results given in
refs.~\cite{Eichten:2021qbm, Eichten:2022vys}, where $\left[{\cal
    M}^2_S(0)\right]_{12}$ was found to be scale dependent and fully
proportional to $m_t^4$. As mentioned in section~\ref{sec:intro}, in
phenomenological analyses of the SI-THDM it is customary to treat
$M_h\approx 125$~GeV as an input, and use the sum rule in
eq.~(\ref{eq:Mhsum}) to constrain one of the masses of the BSM Higgs
bosons. We note that, in principle, a further constraint on the
BSM-Higgs masses could be imposed by requiring that $\left[{\cal
    M}^2_S(0)\right]_{12}=0$, in analogy with the one-loop alignment
condition discussed in refs.~\cite{Degrassi:2023eii,
  Degrassi:2025pqt}. However, eq.~(\ref{eq:MS12}) shows that a
cancellation between the BSM-Higgs and top contributions to the mixing
term is possible only for $\tan\beta<1$.

While the effects of mixing on the scalar masses are formally of
two-loop order and higher, thus beyond the accuracy of the present
calculation, an assessment of their magnitude may still prove useful
(e.g., to check whether their omission in the two-loop calculation of
ref.~\cite{Braathen:2020vwo} was justified). To this purpose, we
compute the eigenvalues $M^2_{h^\prime}$ and $M^2_{H^\prime}$ of the
full scalar mass matrix in eq.~(\ref{eq:MS2p}). More precisely, to
properly account for the momentum dependence of the calculation and
for the possibility of crossing between states, we define
$M_{h^\prime}$ as the mass of the most SM-like eigenstate of the mass
matrix computed at $p^2=0$, and $M_{H^\prime}$ as the mass of the
least SM-like eigenstate of the mass matrix computed at
$p^2=m_H^2$. Note that the degree of ``SM-likeness'' of each
eigenstate is determined by the rotation angle that leads from the
basis $(h,H)$ to the basis $(h^\prime,H^\prime)$. We remark that
$M_{h^\prime}$ and $M_{H^\prime}$ are gauge-dependent quantities,
because both $\left[{\cal M}^2_S(m_H^2)\right]_{11}$ and $\left[{\cal
    M}^2_S(0)\right]_{22}$ are gauge dependent. This is just a
reflection of the fact that $M_{h^\prime}$ and $M_{H^\prime}$ contain
incomplete sets of two-loop effects.

To conclude this section, we comment on the choice of the
renormalization scale $Q$ entering the one-loop quantities in
eqs.~(\ref{eq:Mh1loop})--(\ref{eq:MHp1loop}). In earlier studies of
the SI-THDM it was customary to use the so-called Gildener-Weinberg
scale $Q_{\smallGW}$, defined in ref.~\cite{Gildener:1976ih} as the
scale at which the tree-level potential has a flat direction along $h$
and the one-loop SM-like tadpole $T_h$ vanishes. As we discuss in more
detail in the appendix B, this particular choice of scale is by no
means compulsory; after all, the scale dependence of the individual
terms in eqs.~(\ref{eq:Mh1loop})--(\ref{eq:MHp1loop}) cancels out in
their sum. We also note that refs.~\cite{Eichten:2021qbm,
  Eichten:2022vys} assume that, at the scale $Q_{\smallGW}$, the
minimum conditions for the scalar potential given in
eq.~(\ref{eq:mintree}) remain valid even beyond the tree level. As is
clear from eq.~(\ref{eq:minloop}), this assumption is not correct,
because $T_H$ generally does not vanish at the scale at which $T_h$
vanishes. Similarly, the condition $T_H=0$ imposed in
ref.~\cite{Nhi:2025iob} is an arbitrary constraint on the parameter
space of the model.

\section{Numerical impact of the corrections to the Higgs-boson masses}
\label{sec:numerical}

To illustrate the numerical impact of the corrections to the
Higgs-boson masses in the SI-THDM, we start by adopting a benchmark
scenario similar to the one considered in fig.~1 of
ref.~\cite{Eichten:2022vys}. Once we take into account the minimum
conditions for the scalar potential, the free parameters of the model
can be chosen as three quartic couplings of the $Z_2$-symmetric basis,
namely $\lambda_3$, $\lambda_4$, and $\lambda_5$, plus the ratio of
vevs $\tan\beta$, all interpreted as $\msbar$-renormalized parameters
at some scale $Q$. We treat the one-loop-induced mass of the SM-like
Higgs boson as an input, $M_h=125$~GeV, and we use
eq.~(\ref{eq:Mhsum}) to constrain $\lambda_3$ as a function of
$\lambda_4$ and $\lambda_5$\,, identifying $m_H^2$, $m_A^2$, and
$m_{H^{\pm}}^2$ with the tree-level parts of the expressions for the
respective pole masses in
eqs.~(\ref{eq:MH1loop})--(\ref{eq:MHp1loop}).
We also assume the constraint $\lambda_4=\lambda_5$\,, in which case
$m_A=m_{ H^{\pm}}$\,, see eq.~(\ref{eq:massestree}), and the one-loop
BSM-Higgs contributions to the $\rho$ parameter
vanish~\cite{Toussaint:1978zm}, limiting possible tensions with
precision EW measurements. Finally, we fix $\tan\beta=2$, so that the
BSM-Higgs contributions to the $hH$ mixing term in eq.~(\ref{eq:MS12})
add up with the top-quark contribution.\footnote{This is equivalent to
  the choice $\tan\beta=0.5$ in ref.~\cite{Eichten:2022vys}, because
  in that paper the top quark couples only to $\Phi_1$.} As a result
of these choices, in our benchmark scenario the Higgs-mass spectrum of
the SI-THDM depends on a single quartic coupling of the
$Z_2$-symmetric basis.

\begin{figure}[t]
  \begin{center}
   \includegraphics[width=13cm]{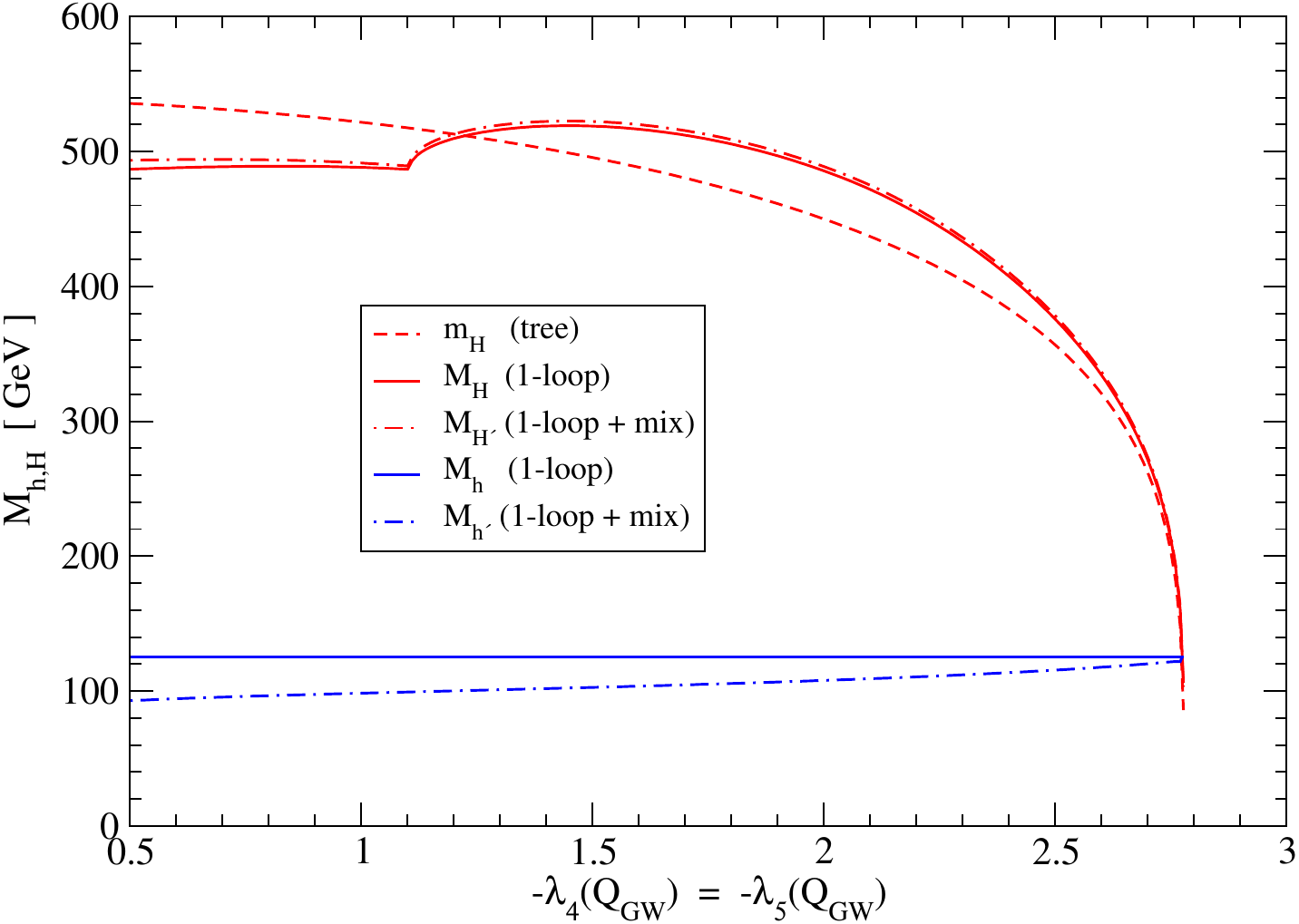}
  \caption{\em Neutral-scalar masses as a function of the quartic
    couplings, in a scenario where $\lambda_4=\lambda_5$, $\lambda_3$
    is fixed by the value of $M_h$, and $\tan\beta=2$. The meaning of
    the lines is explained in the text.}
  \label{fig:MhMHvsL4L5}
\vspace*{-5mm}
\end{center}
\end{figure}

In fig.~\ref{fig:MhMHvsL4L5} we plot the predictions for the neutral
scalar masses in our benchmark scenario as a function of $-\lambda_4$
(or, equivalently, $-\lambda_5$) at different levels of accuracy. The
quartic couplings on the $x$-axis are interpreted as
$\msbar$-renormalized parameters at the scale $Q_{\smallGW}$, for
which an explicit expression is given in the appendix B. In the range
covered by the plot, the values of the quartic couplings correspond to
tree-level masses $m_A$ and $m_{ H^{\pm}}$ that increase between
approximately $174$~GeV and $410$~GeV. The red dashed line in
fig.~\ref{fig:MhMHvsL4L5} represents the tree-level value of the
BSM-Higgs mass $m_H$ as extracted from eq.~(\ref{eq:Mhsum}) for a
given value of $m_A$ and $m_{ H^{\pm}}$: note that it drops to zero
when the sum rule for $M_h$ is saturated by the pseudoscalar and
charged-scalar masses alone. The red and blue solid lines correspond
to the full one-loop predictions for $M_H$ and $M_h$, respectively, as
given in eqs.~(\ref{eq:MH1loop}) and (\ref{eq:Mh1loop}). While the
one-loop prediction for the SM-like Higgs mass is fixed by
construction to $M_h=125$~GeV, we see that the one-loop prediction for
the mass of the BSM Higgs scalar can differ by almost $50$~GeV from
the tree-level mass, i.e., the one-loop corrections are of ${\cal
  O}(10\%)$. This is to be contrasted with the analogous plot in
fig.~1 of ref.~\cite{Eichten:2022vys}, where the tree-level and
one-loop predictions for the $H$ mass are essentially overlapping. We
also note, in our fig.~\ref{fig:MhMHvsL4L5}, a kink in the prediction
for $M_H$ around $\lambda_4,\lambda_5\approx -1.1$\,. This stems from
the thresholds in ${\rm Re}\,\Pi_{HH}(m_H^2)$ for $m_H = 2\,m_A$ and
for $m_H = 2\,m_{ H^{\pm}}$, and would obviously not be reproduced by
a pure effective-potential calculation.

To assess the effects of mixing, which we recall are formally of
two-loop order and higher, we plot the eigenvalues $M_{H^\prime}$ and
$M_{h^\prime}$ of the full mass matrix as the red and blue dot-dashed
lines, respectively. By comparing with the corresponding solid lines,
we see that in this scenario the difference between $M_{H^\prime}$ and
$M_{H}$ amounts to a few GeV, i.e., it is of ${\cal O}(1\%)$. In
contrast, $M_{h^\prime}$ can be lower than $M_{h}$ by more than
$30$~GeV at the lowest considered values of
$|\lambda_4|,|\lambda_5|$. The difference between $M_{h}$ and
$M_{h^\prime}$ then decreases for increasing (absolute) values of the
quartic couplings, due to the strong dependence of the $hH$ mixing
term in the mass matrix, see eq.~(\ref{eq:MS12}), on the BSM-scalar
mass $m_H$, which is larger on the left side of the plot. Again, this
behavior should be contrasted with the one in fig.~1 of
ref.~\cite{Eichten:2022vys}, where the lines corresponding to $M_{h}$
and $M_{h^\prime}$ are essentially overlapping. We also recall that
$M_{H^\prime}$ and $M_{h^\prime}$ are gauge-dependent quantities. The
values shown in fig.~\ref{fig:MhMHvsL4L5} are obtained with $\xi=0$,
while the values obtained with $\xi=1$ can differ by a few tens of MeV
(i.e., negligibly) in the case of $M_{H^\prime}$, but up to $1.5$~GeV
in the case of $M_{h^\prime}$. As mentioned in the previous section,
this is merely a reflection of the fact that $M_{h^\prime}$ and
$M_{H^\prime}$ contain incomplete sets of two-loop corrections.

The non-negligible impact of the mixing effects on $M_{h^\prime}$
suggests that a two-loop calculation along the $h$ direction alone, as
was performed in ref.~\cite{Braathen:2020vwo}, is not sufficient for a
precise prediction of the SM-like Higgs mass in the SI-THDM. On the
other hand, by computing the angle that rotates $(h,H)$ into
$(h^\prime,H^\prime)$, we find that, in this scenario, the couplings
of $h^\prime$ to gauge bosons and to the top quark are modified
w.r.t.~the SM predictions by $\kappa_V\approx 0.99$ and
$\kappa_t\approx 1.06$, respectively. This is well within the
$2\sigma$ bounds from the Run~2 of the
LHC~\cite{ParticleDataGroup:2024cfk}. The impact of the mixing effects
on the couplings of $h^\prime$ to down-type fermions depends on the
form of the Higgs-fermion interactions: in a ``type-I'' THDM, where
all fermions couple only to $\Phi_2$ in the $Z_2$-symmetric basis, we
get $\kappa_{b,\tau,\mu} = \kappa_t$, also within the $2\sigma$
bounds. However, in a ``type-II'' THDM, where the down-type fermions
couple only to $\Phi_1$, we get $\kappa_{b,\tau,\mu} \lesssim 0.74$ in
this scenario, in some tension with the $2\sigma$ bounds at least for
what concerns $\kappa_\tau$.\footnote{Note that the identification
of $h^\prime$ with the observed Higgs boson is anyway precluded by the
predictions for $M_{h^\prime}$.}

\begin{figure}[t]
\begin{center}
  \includegraphics[width=8.3cm]{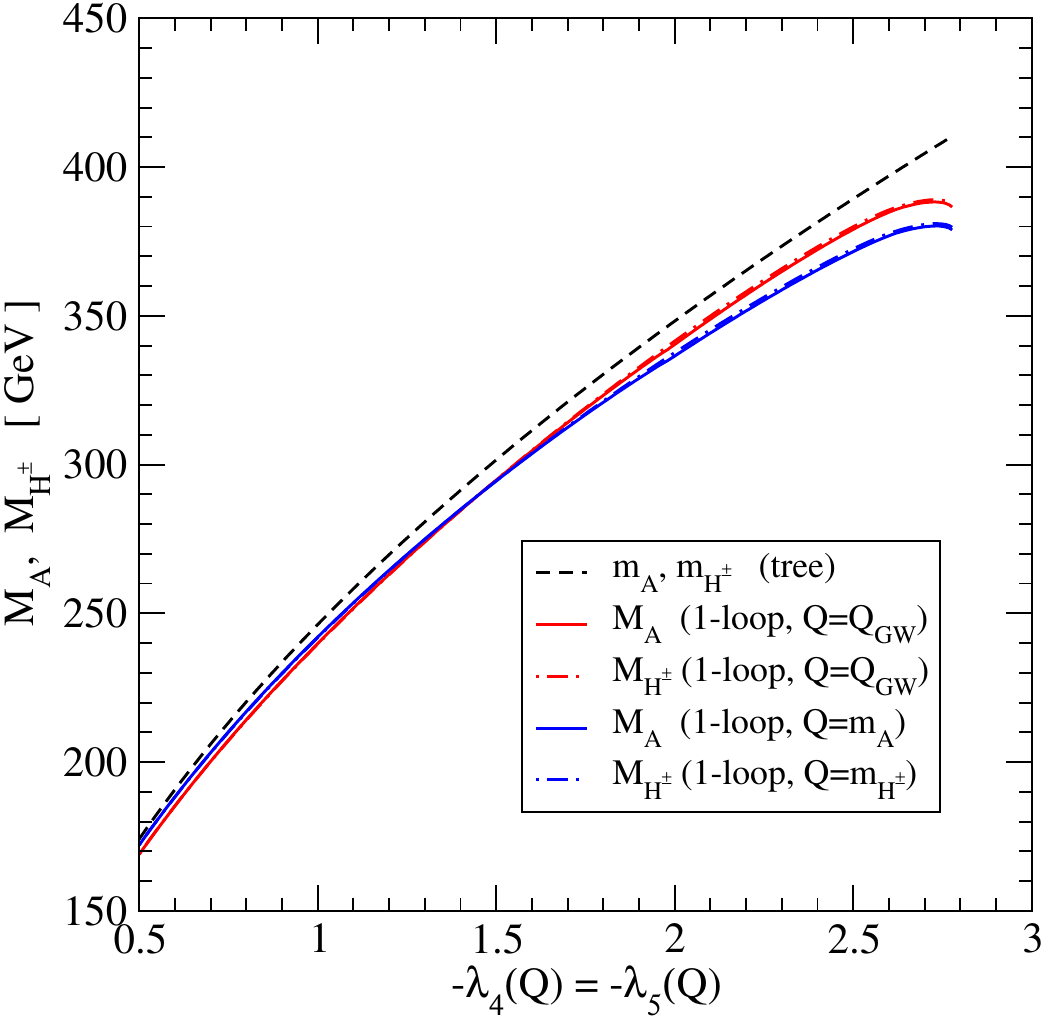}~~~
  \includegraphics[width=8.3cm]{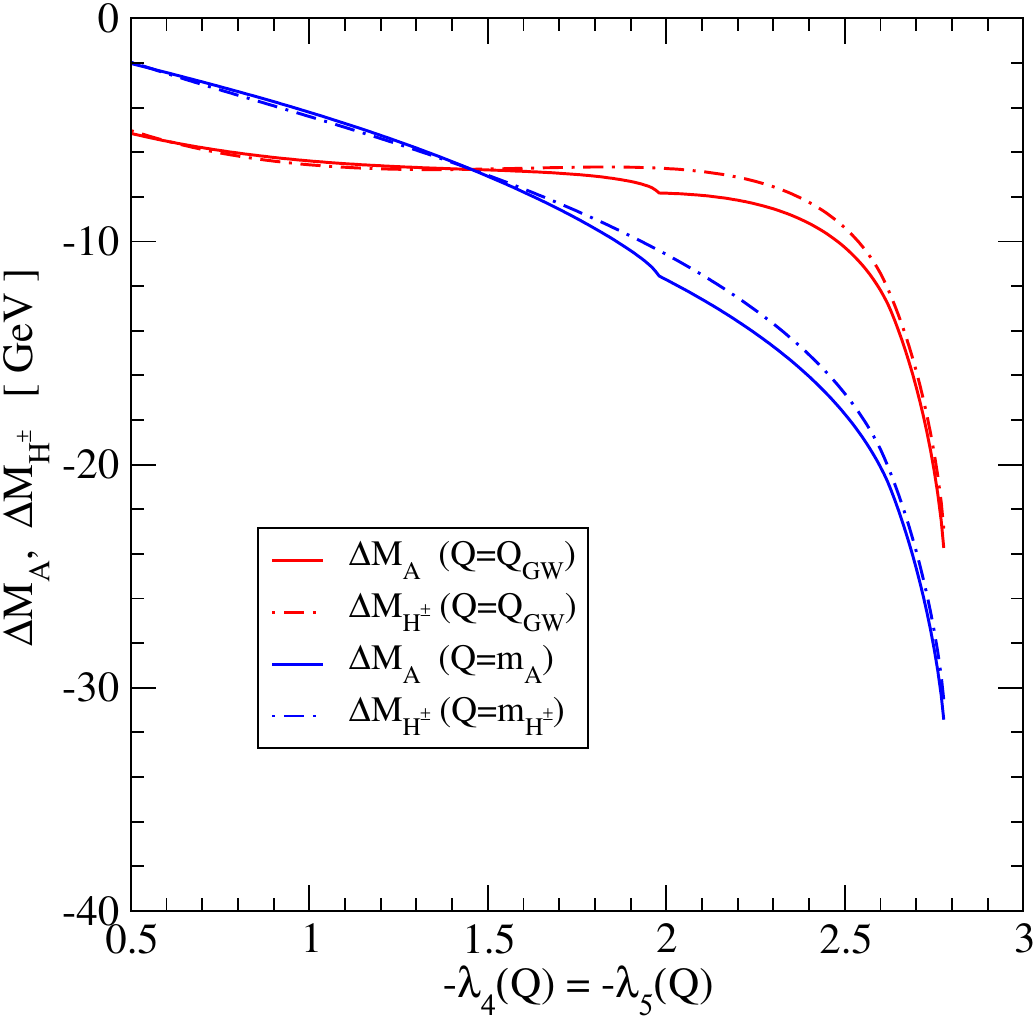}
  \caption{\em Left:~~Pseudoscalar and charged-scalar masses as a
    function of the quartic couplings, in the same scenario as in
    fig.~\ref{fig:MhMHvsL4L5}. ~Right:~~Differences between
    loop-corrected and tree-level masses.  The meaning of the lines is
    explained in the text.}
  \label{fig:MAMHpvsL4L5}
\vspace*{-5mm}
\end{center}
\end{figure}

\vfill
\newpage

We now move on to discussing the impact of the one-loop corrections to
the masses of the pseudoscalar and of the charged scalar. In the left
plot of fig.~\ref{fig:MAMHpvsL4L5} we show the predictions for those
masses as a function of $-\lambda_4$ (or, equivalently, $-\lambda_5$),
in the same scenario as in fig.~\ref{fig:MhMHvsL4L5}. The black dashed
line represents the common tree-level masses $m_A=m_{H^\pm}$, while
the solid and dot-dashed lines correspond to the full one-loop
predictions for $M_A$ and $M_{H^\pm}$, respectively, as given in
eqs.~(\ref{eq:MA1loop}) and (\ref{eq:MHp1loop}). We show two sets of
solid and dot-dashed lines: the red lines are the predictions obtained
assuming that the quartic couplings on the $x$-axis are evaluated at
the scale $Q_\smallGW$ defined in the appendix B, while the blue lines
assume that the couplings are evaluated at a scale equal to the
tree-level masses $m_A$ and $m_{H^\pm}$. We stress that the red and
blue lines in fig.~\ref{fig:MAMHpvsL4L5} map two different regions of
the SI-THDM parameter space, thus their spread should not be
interpreted as a theory uncertainty of the one-loop calculation. To
allow for a better assessment of the size and behavior of the one-loop
corrections, in the right plot of fig.~\ref{fig:MAMHpvsL4L5} we show
the differences between the loop-corrected pseudoscalar and
charged-scalar masses and the corresponding tree-level masses. The
meaning of the color and the style of the different lines is the same
as in the left plot.

Fig.~\ref{fig:MAMHpvsL4L5} shows that, contrary to the finding in
refs.~\cite{Lee:2012jn, Lane:2018ycs, Lane:2019dbc}, the pseudoscalar
and charged-scalar masses do receive one-loop corrections of the order
of a few percent. Comparing the solid and dot-dashed lines in each of
the blue and red sets, we also see a small split between $M_A$ and
$M_{H^\pm}$, which is driven by the corrections controlled by the EW
gauge couplings or by the top Yukawa coupling. In particular, the
small kinks in the solid lines around $\lambda_5\approx -2$ stem from
the threshold in ${\rm Re}\,\Pi_{AA}(m_A^2)$ for $m_A=2\,m_t$, while
${\rm Re}\,\Pi_{H^{\pm}H^{\mp}}(m_{H^{\pm}}^2)$ does not have a
threshold there because, in that case, the fermion loop involves both
top and bottom quarks.  In contrast, the corrections controlled by the
quartic scalar couplings are the same for the pseudoscalar and the
charged scalar, due to an unbroken SO(3) symmetry of the scalar
potential of the SI-THDM that is realized when
$\lambda_4=\lambda_5$~\cite{Pilaftsis:2011ed}. Finally, comparing the
red and blue sets of lines we see that the one-loop corrections to the
pseudoscalar and charged-scalar masses are of the same order of
magnitude independently of whether the $\msbar$-renormalized quartic
couplings entering the tree-level masses are taken as input at
$Q=Q_\smallGW$ or at $Q=m_A,m_{H^\pm}$, although the former choice
yields a somewhat flatter dependence of the corrections on the
couplings before the rapid fall around $\lambda_4,\lambda_5 \approx
-2.5$\,. As discussed in the appendix B, any choice of renormalization
scale for the input parameters that is in the same ballpark as $v$ is
bound to lead to a reasonable perturbative behavior of the loop
corrections.

\bigskip

Deviations from the SM predictions of a few percent in the couplings
of the mass eigenstate $h^\prime$ to fermions and gauge bosons, such
as the ones we found in the scenario of fig.~\ref{fig:MhMHvsL4L5}, are
currently within the LHC bounds, but can in principle be tested at the
HL-LHC~\cite{Cepeda:2019klc}. However, eq.~(\ref{eq:MS12}) shows that
it is even possible to devise an aligned scenario in which the
effective mixing angle between the SM-like and BSM scalar states -- as
defined from the one-loop mass matrix computed at zero external
momentum -- vanishes altogether, so that $h^\prime$ coincides with $h$
and its couplings to fermions and gauge bosons are fully
SM-like. Together with the sum rule for $M_h$ in eq.~(\ref{eq:Mhsum})
and with the condition $m_A=m_{H^\pm}$, which we retain to improve
compatibility with the precision EW measurements, the condition
$\left[{\cal M}^2_S(0)\right]_{12}=0$ fixes all three of the BSM-Higgs
masses $m_H$, $m_A$, and $m_{H^\pm}$. Even in this case we need to
specify a renormalization scale $Q$ for those masses, because in the
calculation of the one-loop corrections we identify them with the
tree-level parts of eqs.~(\ref{eq:MH1loop})--(\ref{eq:MHp1loop}),
which involve the $\msbar$-renormalized quartic couplings. The
remaining free parameter is then $\tan\beta$, and eq.~(\ref{eq:MS12})
shows that a cancellation between the BSM-Higgs and top contributions
to the mixing is possible only when $\tan\beta<1$.

\begin{figure}[t]
  \begin{center}
  \includegraphics[width=8.3cm]{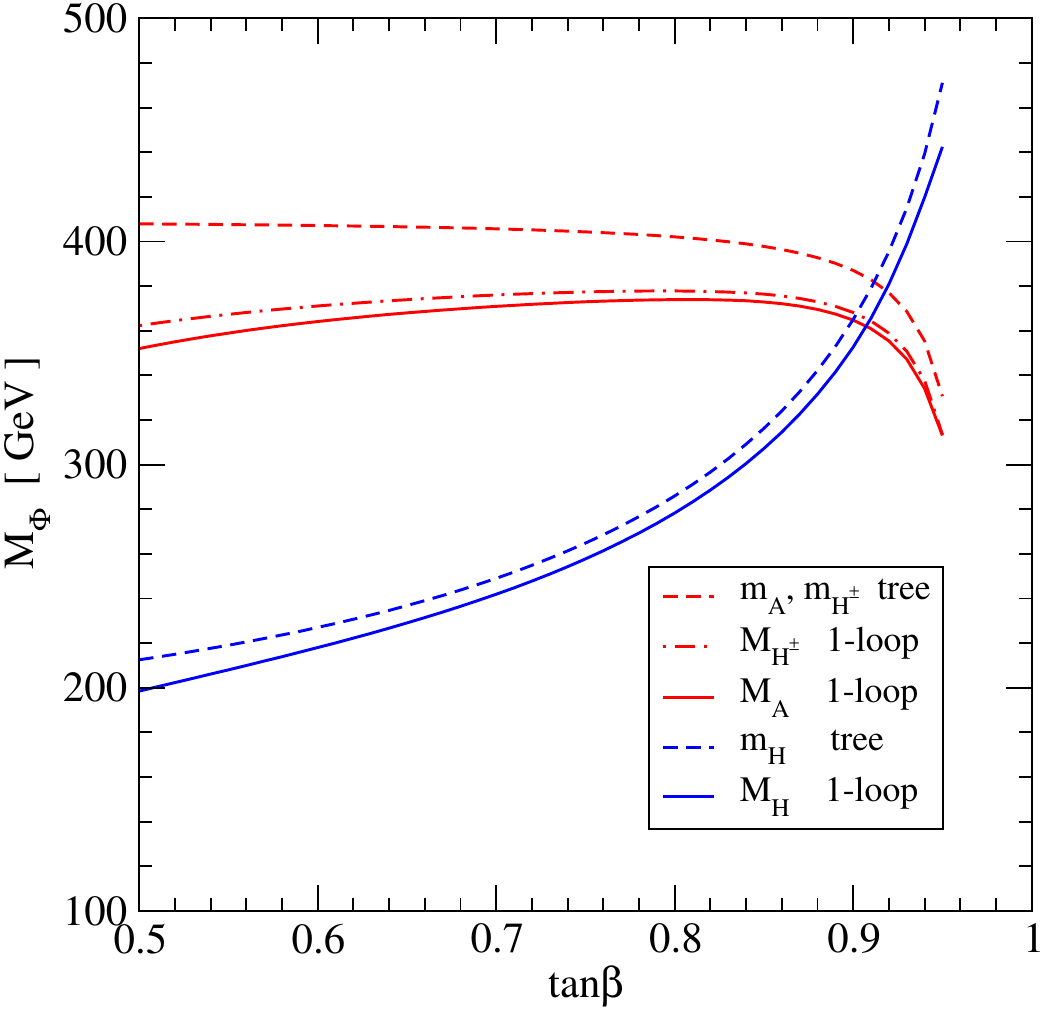}~~~
  \includegraphics[width=8.3cm]{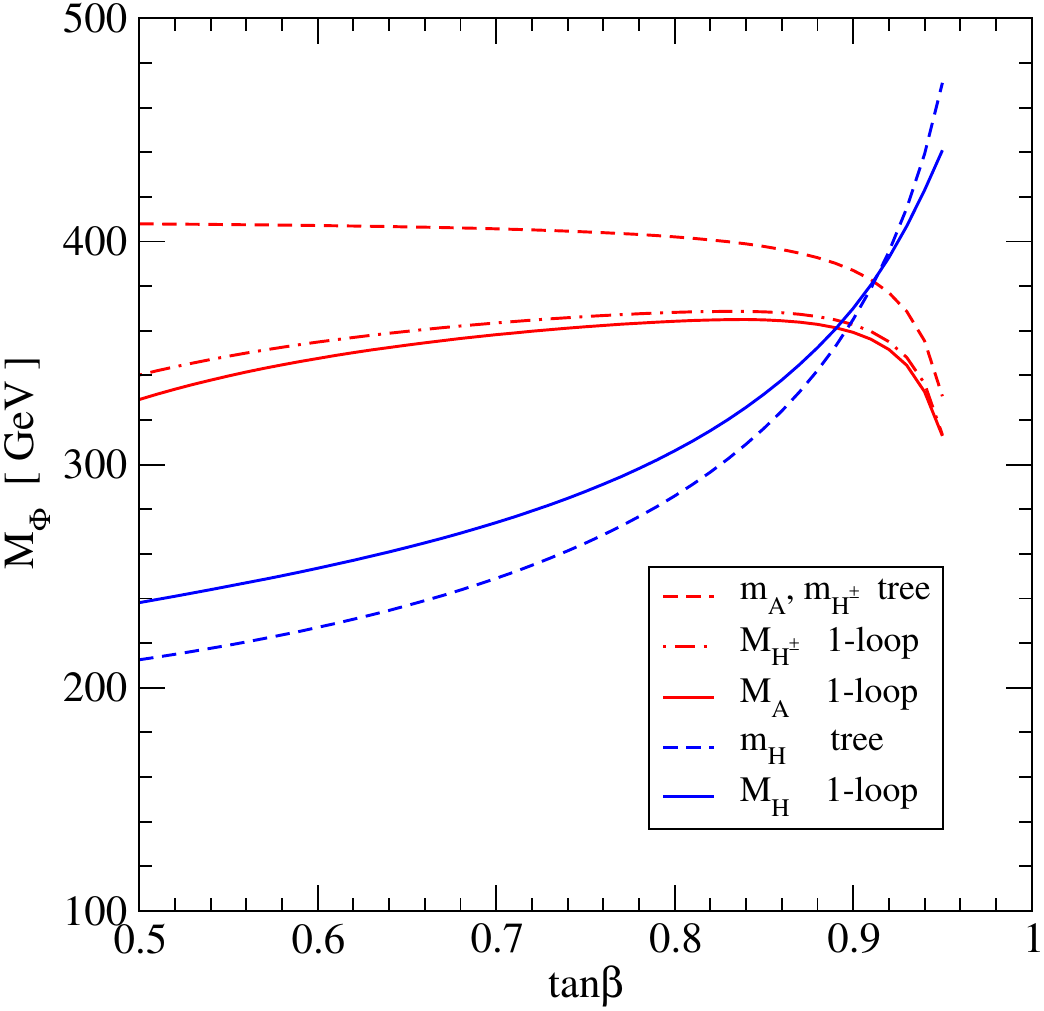}
  \caption{\em Left:~~Masses of the BSM Higgs bosons as a function of
    $\tan\beta$, in an aligned scenario in which $m_A = m_{H^\pm}$,
    $M_h=125$~GeV and $\left[{\cal M}^2_S(0)\right]_{12}=0$, with
    $Q=Q_\smallGW$.  Right:~~Same as the left plot, but with
    $Q=m_A,m_{H^\pm}$\,. The meaning of the lines is explained in the
    text.}
  \label{fig:massesvstanB}
\vspace*{-5mm}
\end{center}
\end{figure}

In the plots of fig.~\ref{fig:massesvstanB} we show the predictions
for the masses of the BSM Higgs bosons as a function of $\tan\beta$ in
this aligned scenario.  The dashed lines represent the common
tree-level masses $m_A=m_{H^\pm}$ (red) and the tree-level mass $m_H$
(blue) as obtained from the combined conditions $M_h=125$~GeV and
$\left[{\cal M}^2_S(0)\right]_{12}=0$ for a given value of
$\tan\beta$. We start from $\tan\beta = 0.5$ to limit the size of the
top Yukawa coupling -- recalling that $y_t = \sqrt2\,m_t/(v\,s_\beta)$
-- and we find solutions up to $\tan\beta\approx0.95$. The solid and
dot-dashed red lines represent the one-loop predictions for $M_A$ and
$M_{H^{\pm}}$, respectively, as given in eqs.~(\ref{eq:MA1loop}) and
(\ref{eq:MHp1loop}), while the solid blue line represents the one-loop
prediction for $M_H$ as given in eq.~(\ref{eq:MH1loop}). The left plot
is obtained with $Q=Q_\smallGW$, and the right plot is obtained with
$Q=m_A,m_{H^\pm}$. We stress again that the two choices of scale
correspond to two different slices of the SI-THDM parameter space,
thus the differences between the two plots should not be interpreted
as a theory uncertainty of the one-loop calculation.

Fig.~\ref{fig:massesvstanB} shows that in this aligned scenario all of
the BSM-Higgs masses are relatively light, and that the charged-scalar
and pseudoscalar masses are constrained in a band between
approximately $310$~GeV and $370$~GeV (in contrast, the mass of the
BSM neutral scalar ranges between approximately $200$ GeV and $440$
GeV). In this scenario the effects of the radiative corrections on the
BSM-Higgs masses are of ${\cal O}(10\%)$, and the comparison between
the left and right plots shows that the choice $Q=Q_\smallGW$ leads to
somewhat smaller corrections than the choice $Q=m_A,m_{H^\pm}$.

\bigskip

We conclude this section with a comment on the use of the conditions
$M_h=125$~GeV and $\left[{\cal M}^2_S(0)\right]_{12}=0$ to constrain
some of the BSM-Higgs masses. While this procedure proves convenient
for defining benchmark scenarios and for comparing with earlier
analyses in the literature, the discussion of
figs.~\ref{fig:MAMHpvsL4L5} and~\ref{fig:massesvstanB} above shows
that it suffers from intrinsic ambiguities. Indeed, since both $M_h$
and $\left[{\cal M}^2_S(0)\right]_{12}$ are purely one-loop-induced
quantities, see eqs.~(\ref{eq:Mhsum}) and (\ref{eq:MS12}), the precise
definitions of the BSM-Higgs masses entering their expressions amount
to higher-order effects. Thus, we are free to identify those masses
with the tree-level parts of
eqs.~(\ref{eq:MH1loop})--(\ref{eq:MHp1loop}), but then, as shown in
figs.~\ref{fig:MAMHpvsL4L5} and~\ref{fig:massesvstanB}, different
choices of scale for the $\msbar$-renormalized quartic couplings
entering the tree-level masses correspond to different regions of the
parameter space. Note that we could even choose to identify the
BSM-Higgs masses entering eqs.~(\ref{eq:Mhsum}) and (\ref{eq:MS12})
directly with the pole masses, which would correspond to yet another
region of the parameter space. These ambiguities could in principle be
resolved by computing $M_h$ and $\left[{\cal M}^2_S(0)\right]_{12}$ at
the two-loop level, although in that case the extraction of the
relevant relations between the masses would be nontrivial and likely
to require numerical methods. This said, we stress that there is no
ambiguity in the one-loop determination of the BSM-Higgs masses via
eqs.~(\ref{eq:MH1loop})--(\ref{eq:MHp1loop}) when all three of the
$\msbar$-renormalized couplings $\lambda_3$, $\lambda_4$, and
$\lambda_5$ are directly treated as input at a given renormalization
scale.

\section{Conclusions}
\label{sec:conclusions}

The most attractive features of the SI-THDM are a constrained mass
spectrum for the BSM Higgs bosons, which must all be relatively light,
and tree-level alignment, so that one of the neutral scalars has
SM-like couplings to gauge bosons and fermions. However, most of the
earlier calculations of the Higgs-mass spectrum of the SI-THDM, which
relied on the effective-potential approach, were plagued by an
incorrect assumption for the field-dependent masses~\cite{Lee:2012jn,
  Lane:2018ycs, Lane:2019dbc, Eichten:2021qbm, Eichten:2022vys} and/or
by an incorrect treatment of the minimum conditions for the scalar
potential~\cite{Eichten:2021qbm,Eichten:2022vys,Nhi:2025iob}. In this
paper we revisited the one-loop calculation of the Higgs-mass spectrum
of the SI-THDM, but we relied on a direct calculation of the relevant
Feynman diagrams. This allowed us to bypass the complications related
to the definition of the field-dependent masses, and also to account
for the external-momentum effects, which are necessary to ensure the
scale and gauge independence of the pole masses of the BSM Higgs
bosons. Explicit formulas for all of the one-loop quantities relevant
to our calculation are provided in the appendix A.

Exploring the numerical implications of our calculation, we showed
that, in contrast with the findings of the earlier papers, the
one-loop corrections to the BSM-Higgs masses can be of ${\cal
  O}(10\%)$. Furthermore, we found that the one-loop corrections can
induce a non-negligible mixing between the SM-like and BSM states in
the neutral-scalar sector. This induces a shift in the mass of the
SM-like state that is formally of two-loop order, but was not taken
into account in the two-loop calculations of
refs.~\cite{Braathen:2020vwo, Eichten:2022vys}. We then considered an
aligned scenario in which the condition of vanishing mixing induces a
further constraint on the BSM-Higgs masses. Finally, we discussed how
the definitions of benchmark scenarios for the SI-THDM can depend on
the choice of renormalization scale, and elucidated the role of the
Gildener-Weinberg scale introduced in ref.~\cite{Gildener:1976ih} (see
also the appendix B).

\bigskip

The work presented in this paper could of course be extended in
several directions. On the calculation side, our full one-loop results
for the neutral-scalar mass matrix could be combined with the two-loop
results for $M_h$ in ref.~\cite{Braathen:2020vwo}, so that the mixing
effects are properly taken into account in the latter. The two-loop
calculation could in turn be extended to the full mass matrix for the
neutral scalars, thus correcting the results of
ref.~\cite{Eichten:2022vys}, and even to the pseudoscalar and
charged-scalar masses. On the phenomenology side, the constraints on
the parameter space of the SI-THDM that arise from the measured
properties of the $125$-GeV Higgs boson and from the searches for BSM
Higgs bosons at the LHC could be reassessed in view of the correct
calculation of masses and mixing. We leave all of this for future
work, hoping that the results presented in this paper will help the
collective effort to use the Higgs sector as a probe of what lies
beyond the SM.

\section*{Acknowledgments}

We thank G.~Degrassi, M.~Goodsell, and E.~Senaha for useful
discussions.

\vfill
\newpage

\section*{Appendix A: One-loop tadpoles and self-energies in the SI-THDM}
\setcounter{equation}{0}
\renewcommand{\theequation}{A\arabic{equation}}

In this appendix we list explicit formulas for the one-loop quantities
that are relevant to our calculation, including also the counterterm
to the relation between $v$ and $G_F$. The tadpoles and self-energies
for the physical scalar fields $(h,H,A,H^\pm)$ in a generic $R_\xi$
gauge are

\bea
16\pi^2\, v\,T_h &=&\mHq\,A_0(\mHq)\,+\,\mAq\,A_0(\mAq)
  \,+\,2\,\mHpq\,A_0(\mHpq) \,-\, 4\,N_c\,m_t^2\,A_0(m_t^2)\nn\\[2mm]
  &+& \mZq\,\left[3\,A_0(\mZq) + 2\,\mZq\right] \,+\,
  2\,\mWq\,\left[3\,A_0(\mWq) + 2\,\mWq\right]~,\\[5mm]
16\pi^2\, v\,T_H &=&
\mHq\,\cot2\beta\,\biggr[
  3\,A_0(\mHq) \,+\, A_0(\mAq) \,+\, 2\, A_0(\mHpq)\biggr]
-\, 4\,N_c\,m_t^2\,\cot\beta\,A_0(m_t^2)
~,\nn\\\\
\label{eq:Pihh}
16\pi^2\,  v^2\,\Pi_{hh}(p^2) &=&
\mHq\,F(p^2,\mHq)\,+\,\mAq\,F(p^2,\mAq)\,+\,
2\,\mHpq\,F(p^2,\mHpq)\nn\\[2mm]
&-& \!2\,N_c\, m_t^2\, 
\biggr[ 2\,F(p^2,m_t^2) + p^2\,B_0(p^2,m_t^2,m_t^2)\biggr]\nn\\[2mm]
&+& f_1(p^2,\mZq) \,+\,2\,f_1(p^2,\mWq)~,\\[5mm]
16\pi^2\,  v^2\, \Pi_{hH}(p^2) &=&
\mHq\,\cot2\beta\,\biggr[
  3\,F(p^2,\mHq)\,+\,F(p^2,\mAq)\,+\,2\,F(p^2,\mHpq)\biggr]\nn\\[2mm]
&-& 2\,N_c\,m_t^2\, \cot\beta\,
\biggr[ 2\,F(p^2,m_t^2) + p^2\,B_0(p^2,m_t^2,m_t^2) \biggr]
~,\\[5mm]
16\pi^2\,  v^2\, \Pi_{HH}(p^2) &=&
2\,\mHq\,\cot^22\beta\,
\biggr[3\,A_0(\mHq)+A_0(\mAq)+2\,A_0(\mHpq)\biggr]
\,-\,4\,m_{H}^4\,B_0(p^2,\mHq,0)\nn\\[2mm]
&-&2\,m_{H}^4\,\cot^22\beta\,\biggr[
  9\,B_0(p^2,\mHq,\mHq) + B_0(p^2,\mAq,\mAq) + 2\,B_0(p^2,\mHpq,\mHpq)\biggr]
\nn\\[2mm]
&-& 2\,N_c\,m_t^2\, \cot^2\beta\,
\biggr[ 2\,F(p^2,m_t^2) + p^2\,B_0(p^2,m_t^2,m_t^2) \biggr]\nn\\[2mm]
&+&f_2(p^2,\mZq,\mAq,\mHq)\,+\,2\,f_2(p^2,\mWq,\mHpq,\mHq)~,
  \\[5mm]
16\pi^2\,  v^2\, \Pi_{AA}(p^2) &=&
2\,\mHq\,\cot^22\beta\,
\biggr[A_0(\mHq)+3\,A_0(\mAq)+2\,A_0(\mHpq)\biggr]
\nn\\[2mm]
&-&4\,m_A^4\,B_0(p^2,\mAq,0)
  \,-\,4\,m_H^4\,\cot^22\beta\,B_0(p^2,\mAq,\mHq)\nn\\[2mm]
&-&2\,N_c\,m_t^2\, \cot^2\beta\,
  \biggr[ 2\,A_0(m_t^2) + p^2 \,B_0(p^2,m_t^2,m_t^2) \biggr]\nn\\[2mm]
  &+&f_2(p^2,\mZq,\mHq,\mAq) \,+\, 2\,f_2(p^2,\mWq,\mHpq,\mAq)~,
  \\[5mm]
16\pi^2\,  v^2\, \Pi_{H^+H^-}(p^2) &=&
2\,\mHq\,\cot^22\beta\,
\biggr[A_0(\mHq)+A_0(\mAq)+4\,A_0(\mHpq)\biggr]
\nn\\[2mm]
&-&4\,m_{H^\pm}^4\,B_0(p^2,\mHpq,0)
  \,-\,4\,m_H^4\,\cot^22\beta\,B_0(p^2,\mHpq,\mHq)\nn\\[2mm]
&-&2\,N_c\,m_t^2\, \cot^2\beta\,
  \biggr[A_0(m_t^2) + (p^2-m_t^2)\,B_0(p^2,m_t^2,0) \biggr]\nn\\[2mm]
  &+&\!\!\!  f_2(p^2,\mWq,\mHq,\mHpq) +f_2(p^2,\mWq,\mAq,\mHpq)
  +c^2_{2\theta_W}f_2(p^2,\mZq,\mHpq,\mHpq)\nn\\[2mm]
  &+&  \!s^2_{2\theta_W}\,\mZq\,\biggr[\,
    2\,(\xi-1)\,p^2\,+\,(3-2\,\xi)\,A_0(\mHpq)\,
    +\,\frac{\,\mHpq}{\mZq}\,A_0(\xi\,\mZq)\nn\\[2mm]
    &&~~~~~~~~~~~~+\,(3-\xi)\,(p^2+\mHpq)\,B_0(p^2,\mHpq,0)\,\biggr]~,
\label{eq:PiHpHm}
\eea
where $\theta_W$ is the weak mixing angle defined by $c_{\theta_W} =
m_W/m_Z$, and $A_0(m^2)$ and $B_0(p^2,m_1^2,m_2^2)$ are
Passarino-Veltman functions.  In particular, we can write:
\beq
\label{eq:A0B0}
A_0(m^2) ~=~ m^2\left(\ln\frac{m^2}{Q^2}-1\right)~,~~~~~
B_0(0,m_1^2,m_2^2) ~=~ -\frac{A_0(m_1^2)- A_0(m_2^2)}{m_1^2-m_2^2}~,
\eeq
while we use the code {\tt LoopTools}~\cite{Hahn:1998yk} to account
for the full $p^2$-dependence of $B_0$. In
eqs.~(\ref{eq:Pihh})--(\ref{eq:PiHpHm}) we also introduced the
auxiliary functions
\bea
F(p^2,m^2) &=& A_0(m^2) \,-\,2\,m^2\,B_0(p^2,m^2,m^2)~,\\[5mm]
f_1(p^2,m_V^2) &=& 3\,m_V^2\,\left[A_0(m_V^2)+2\,m_V^2\right]
\,-\,p^2\,\left[A_0(m_V^2)-A_0(\xi m_V^2)\right]\nn\\[2mm]
  &-&\frac12\,(p^4-4\,p^2\,m_V^2+12\,m_V^4)\,B_0(p^2,m_V^2,m_V^2)
\,+\,\frac{p^4}2\,B_0(p^2,\xi m_V^2,\xi m_V^2)~,~~~~~~\\[5mm]
f_2(p^2,m_V^2,m_1^2,m_2^2) &=& 2\,m_V^4+(2\,m_V^2+m_1^2)\,A_0(m_V^2)
+m_V^2\,A_0(m_1^2)-p^2\left[A_0(m_V^2)-A_0(\xi m_V^2)\right]\nn\\[2mm]
&-&\left[m_1^4+(p^2-m_V^2)^2-2\,m_1^2(p^2+m_V^2)\right]\,B_0(p^2,m_1^2,m_V^2)
\nn\\[2mm]
&+&(p^2-m_2^2)\,(p^2+m_2^2-2\,m_1^2)\,B_0(p^2,m_1^2,\xi m_V^2)~.
\eea

The $\msbar$-renormalized vev $v$ is connected with the Fermi constant
$G_F$ by the relation $(\sqrt2 G_F)^{-1} = v^2 + \delta v^2$, with
\beq
\label{eq:dv2}
\delta v^2 ~=~ v^2\,\left[\frac{\Pi_{WW}(0)}{\mWq} - \delta_{\rm VB}\right]~,
\eeq
where $\Pi_{WW}(0)$ is the finite part of the one-loop self-energy of
the $W$ boson at vanishing momentum, and $\delta_{\rm VB}$ contains
the one-loop contributions to muon decay from vertex and box
diagrams. The SM contributions to $\Pi_{WW}(0)$ and $\delta_{\rm VB}$
were long ago computed in the Feynman gauge ($\xi=1$) in
refs.~\cite{Marciano:1980pb} and \cite{Sirlin:1985ux}, respectively.
Supplementing $\Pi_{WW}(0)$ with the BSM contributions,\footnote{There
are no BSM contributions to $\delta_{\rm VB}$ under our approximations
for the Yukawa couplings.} and setting the tree-level mass of $h$ to
zero in the SM part, we get:
\bea
\label{eq:PiWW0}
\frac{16\pi^2}{g^2}\left.\Pi_{WW}(0)\right|_{\xi=1} &=&
\mWq\biggr[\frac{27-34s_{\theta_W}^2}{8\,c_{\theta_W}^2} \,+\,
  \left(\frac{17}{s_{\theta_W}^2}+8\,s_{\theta_W}^2-29
  \right)\frac{\ln c_{\theta_W}^2}{4c_{\theta_W}^2}
  \,+\,\left(\frac{s_{\theta_W}^2}{c_{\theta_W}^2}
  -1\right)\,\ln\frac{\mWq}{Q^2}\biggr]\nn\\[2mm]
&&-\frac{N_c}{2}\,m_t^2\left(\ln\frac{m_t^2}{Q^2}-\frac12\right)
\,+\,\wt{B}_{22}(0,\mHq,\mHpq)\,+\,\wt{B}_{22}(0,\mAq,\mHpq)~,\\[5mm]
\label{eq:dVB}
\frac{16\pi^2}{g^2}\left.\delta_{\rm VB}\right|_{\xi=1} &=&
6 \,+\,\left(\frac72-6\,s_{\theta_W}^2\right)
\frac{\ln c_{\theta_W}^2}{s_{\theta_W}^2}\,-\,4\,\ln\frac{\mZq}{Q^2}~,
\eea
with
\beq
\widetilde B_{22}(0,m_1^2,m_2^2)~=~
\frac12\,\left(\frac{m_1^2+m_2^2}{4}-\frac{m_1^2\,m_2^2}{2\,(m_1^2-m_2^2)}
\,\ln\frac{m_1^2}{m_2^2}\right)~.
\eeq

The $\xi$-dependent parts of $\Pi_{WW}(0)$ and $\delta_{\rm VB}$ were
computed in ref.~\cite{Degrassi:1992ff}. For the combination that is
relevant to our calculation, we get\,\footnote{We thank G.~Degrassi
for clarifications on this point.}
\beq
\delta v^2 ~=~ \left.\delta v^2\right|_{\xi=1}\,+\,\frac{1}{16\pi^2}
\,(1-\xi)\,\biggr[\mZq\,B_0(0,\mZq,\xi\mZq)\,
  +\,2\,\mWq\,B_0(0,\mWq,\xi\mWq)\biggr]~.
\eeq

Combining the $\xi$-dependent formulas for self-energies and $\delta
v$, we can check that the pole masses in
eqs.~(\ref{eq:Mh1loop})--(\ref{eq:MHp1loop}) are indeed
gauge independent. In the case of the charged-scalar mass, this
requires also the identity $B_0(m^2,m^2,0) = 2-\ln(m^2/Q^2)$\,.

\vfill
\newpage
\section*{Appendix B: On the Gildener-Weinberg scale}
\setcounter{equation}{0}
\renewcommand{\theequation}{B\arabic{equation}}

To elucidate the role of the so-called Gildener-Weinberg scale
introduced in ref.~\cite{Gildener:1976ih}, it is convenient to rotate
the scalar potential of eq.~(\ref{eq:Vstd}) from the $Z_2$-symmetric
basis $(\Phi_1,\Phi_2)$ to the Higgs basis
$(\Phi_\smallSM,\Phi_\smallBSM)$:
\bea
V_0 &=& \frac{\Lambda_1}{2}\,\left(\Phi_\smallSM^\dagger \Phi_\smallSM\right)^2
\,+~ \frac{\Lambda_2}{2}\,\left(\Phi_\smallBSM^\dagger \Phi_\smallBSM\right)^2
\nn\\[2mm]
&&
+~ \Lambda_3\,\left(\Phi_\smallSM^\dagger \Phi_\smallSM\right)
\left(\Phi_\smallBSM^\dagger \Phi_\smallBSM\right)
+~ \Lambda_4\,\left(\Phi_\smallSM^\dagger \Phi_\smallBSM\right)
\left(\Phi_\smallBSM^\dagger \Phi_\smallSM\right)\nn\\[2mm]
&&
+~ \left[\,
\frac{\Lambda_5}{2}\,\left(\Phi_\smallSM^\dagger \Phi_\smallBSM\right)^2
+\, \left( \Lambda_6 \,\Phi_\smallSM^\dagger \Phi_\smallSM
\,+\,\Lambda_7\,\Phi_\smallBSM^\dagger \Phi_\smallBSM \right)
\,
\Phi_\smallSM^\dagger \Phi_\smallBSM ~+~{\rm h.c.}\right]~.
\label{eq:Vhb}
\eea
In this basis the minimum conditions for the effective potential take
on a simple form:
\beq
\label{eq:minloophb}
\Lambda_1 ~=\, - \frac{2}{v^3}\,T_h~,~~~~~~~~~
\Lambda_6 ~=\, - \frac{2}{v^3}\,T_H~,
\eeq
where we made use of the fact that $T_\varphi \,=\, d \Delta
V/d\varphi\,|_{\rm min}\,$.  Eq.~(\ref{eq:minloophb}) above is
equivalent to eq.~(\ref{eq:minloop}), in view of the relations
\beq
\label{eq:L1L6}
\Lambda_1 ~=~ \lambda_1 \,\cbe^4 \,+\, \lambda_2\,\sbe^4
+ \frac12\,\lambda_{345}\,\sdbe^2~,~~~~~~~~
\Lambda_6 ~=~ -\frac12\,\sdbe \,
\left(\lambda_1\,\cbe^2-\lambda_2\,\sbe^2-\lambda_{345}\,\cdbe\right)~.
\eeq
For a choice of scale $Q_{\smallGW}$ such that
$\Lambda_1(Q_{\smallGW})=0$, so that the tree-level potential has a
flat direction along $h$, the first minimum condition in
eq.~(\ref{eq:minloophb}) implies $T_h=0$. We stress again that this
choice does not have any special implication for the second condition
in eq.~(\ref{eq:minloophb}), contrary to what was assumed in
refs.~\cite{Eichten:2021qbm, Eichten:2022vys}. Similarly, the
requirement in ref.~\cite{Nhi:2025iob} that
$\Lambda_6(Q_{\smallGW})=0$, which implies $T_H=0$ via the second
minimum condition in eq.~(\ref{eq:minloophb}), should be interpreted
as an arbitrary constraint on the parameter space of the model.

\bigskip

To work out the implications of the condition $T_h=0$, we follow the
approach of ref.~\cite{Gildener:1976ih} and write the one-loop part of
the effective potential along the $h$ direction in the form:
\beq
\Delta V ~=~ (v+h)^4\,\left[A + B\,\ln\frac{(v+h)^2}{Q^2}\right]~.
\eeq
This is possible because, in scale-invariant models where the scalar
potential does not include explicit mass parameters, all of the
$h$-dependent squared masses entering the effective potential have
the form $\wt \Lambda\,(v+h)^2$, where by $\wt \Lambda$ we denote
generic combinations of quartic scalar couplings $\Lambda_i$, squared
gauge couplings $g^2$ and $g^{\prime\,2}$, or the squared top coupling
$y_t^2$. Working in the Landau gauge ($\xi=0$) and in the $\msbar$
renormalization scheme, the dimensionless parameters $A$ and $B$ read
\beq
\label{eq:AB}
A ~=~ \sum_\varphi \,\frac{\alpha_\varphi\,m_\varphi^4}{64\pi^2v^4}\,\left(\ln\frac{m_\varphi^2}{v^2}
-k_\varphi\right)~,~~~~~~~
B ~=~ \sum_\varphi \,\frac{\alpha_\varphi\,m_\varphi^4}{64\pi^2v^4}~,
\eeq
where $\alpha_\varphi = (1,1,2,6,3,-4\,N_c)$ for $\varphi =
(H,A,H^\pm,W,Z,t)$, $m^2_\varphi$ are the squared masses at the
minimum of the potential, and $k_\varphi$ is equal to $5/6$ for the
gauge bosons and to $3/2$ for the BSM Higgs bosons and the top
quark. Note that both A and B are of ${\cal O}(\wt
\Lambda^2/16\pi^2)$, and that $M_h^2$ as given in eq.~(\ref{eq:Mhsum})
is equal to $8\,v^2 B$. For $Q=Q_{\smallGW}$\,, the condition $T_h = d
\Delta V/dh\,|_{\rm min} = 0$ corresponds to
\beq
\label{eq:QGW}
\ln\frac{v^2}{Q_{\smallGW}^2} ~=~  - \frac12 -\frac{A}{B}~.
\eeq
Since $A$ and $B$ are of the same order in the couplings $\wt \Lambda$
and in the loops, the logarithm in eq.~(\ref{eq:QGW}) is generically
of ${\cal O}(1)$, thus the perturbative expansion should be
safe. Moreover, eq.~(\ref{eq:QGW}) can be interpreted as a Higgs vev
of ${\cal O}(Q_{\smallGW})$ being generated via ``dimensional
transmutation'':
\beq
\label{eq:dimtrans}
v ~=~ e^{- \left(\frac14 +\frac{A}{2B}\right)}\, Q_{\smallGW}~.
\eeq

We now discuss the implications of choosing a different scale $Q$ for
which $\Lambda_1\neq0$. In that case the minimum condition for the
effective potential along the $h$ direction corresponds to:
\beq
\label{eq:QnoGW}
\ln\frac{v^2}{Q^2} ~=~ - \frac12 -\frac{A}{B}
- \frac{\Lambda_1}{8B}~.
\eeq
If we considered $\Lambda_1$ to be of ${\cal O}(\wt \Lambda)$ like the
other couplings in the model, the third term on the r.h.s.~of
eq.~(\ref{eq:QnoGW}) would be of ${\cal O}(16\pi^2/\wt \Lambda)$,
leading to a large logarithm and a breakdown of the perturbative
expansion.\footnote{Note that in ref.~\cite{Gildener:1976ih} the role
  of our $\wt \Lambda$ is played by $e^2$. Curiously, the discussion
  after eq.~(3.24) of that paper seems indeed to suggest that the
  perturbative expansion would break down if we chose a scale
  different from $Q_\smallGW$.} However, this cannot happen as long as
the scales $Q$ and $Q_\smallGW$ are not very far from each other, so
that $\ln(Q/Q_\smallGW)$ remains of ${\cal O}(1)$. In that case, since
$Q_{\smallGW}$ is defined by the condition
$\Lambda_1(Q_{\smallGW})=0$, the renormalization-group evolution to
$Q$ can only induce a value of ${\cal O}(\wt\Lambda^2/16\pi^2)$ for
$\Lambda_1$. Thus, even the third term on the r.h.s.~of
eq.~(\ref{eq:QnoGW}) is in fact of ${\cal O}(1)$, and the perturbative
expansion remains safe.

\bigskip

In summary, we stress again that in the calculation of the Higgs-mass
spectrum of the SI-THDM the choice $Q=Q_{\smallGW}$ is by no means
compulsory. Any scale choice in the same ballpark as the masses of the
particles involved in the loops -- or, equivalently, in the same
ballpark as $v$ -- will lead to a reasonable perturbative expansion,
and anyway the predictions for physical quantities must be independent
of the scale, as we indeed find in
eqs.~(\ref{eq:Mh1loop})--(\ref{eq:MHp1loop}).

\vfill
\newpage

\bibliographystyle{utphys}
\bibliography{CSI}

\providecommand{\href}[2]{#2}\begingroup\raggedright\begin{thebibliography}{10}

\bibitem{CMS:2012qbp}
{\bf CMS} Collaboration, S.~Chatrchyan {\em et al.}, {\em {Observation of a New
  Boson at a Mass of 125 GeV with the CMS Experiment at the LHC}}.
  \href{http://dx.doi.org/10.1016/j.physletb.2012.08.021}{Phys. Lett. B {\bf
  716} (2012)  30--61}, \href{http://arxiv.org/abs/1207.7235}{{\tt
  arXiv:1207.7235 [hep-ex]}}.

\bibitem{ATLAS:2012yve}
{\bf ATLAS} Collaboration, G.~Aad {\em et al.}, {\em {Observation of a new
  particle in the search for the Standard Model Higgs boson with the ATLAS
  detector at the LHC}}.
  \href{http://dx.doi.org/10.1016/j.physletb.2012.08.020}{Phys. Lett. B {\bf
  716} (2012)  1--29}, \href{http://arxiv.org/abs/1207.7214}{{\tt
  arXiv:1207.7214 [hep-ex]}}.

\bibitem{ParticleDataGroup:2024cfk}
{\bf Particle Data Group} Collaboration, S.~Navas {\em et al.}, {\em {Review of
  particle physics}}.
  \href{http://dx.doi.org/10.1103/PhysRevD.110.030001}{Phys. Rev. D {\bf 110}
  (2024) no.~3, 030001}.

\bibitem{Gunion:1989we}
J.~F. Gunion, H.~E. Haber, G.~L. Kane, and S.~Dawson, {\em {The Higgs Hunter's
  Guide}}. Front. Phys. {\bf 80} (2000)  .

\bibitem{Aoki:2009ha}
M.~Aoki, S.~Kanemura, K.~Tsumura, and K.~Yagyu, {\em {Models of Yukawa
  interaction in the two Higgs doublet model, and their collider
  phenomenology}}. \href{http://dx.doi.org/10.1103/PhysRevD.80.015017}{Phys.
  Rev. D {\bf 80} (2009)  015017}, \href{http://arxiv.org/abs/0902.4665}{{\tt
  arXiv:0902.4665 [hep-ph]}}.

\bibitem{Branco:2011iw}
G.~C. Branco, P.~M. Ferreira, L.~Lavoura, M.~N. Rebelo, M.~Sher, and J.~P.
  Silva, {\em {Theory and phenomenology of two-Higgs-doublet models}}.
  \href{http://dx.doi.org/10.1016/j.physrep.2012.02.002}{Phys. Rept. {\bf 516}
  (2012)  1--102}, \href{http://arxiv.org/abs/1106.0034}{{\tt arXiv:1106.0034
  [hep-ph]}}.

\bibitem{Gunion:2002zf}
J.~F. Gunion and H.~E. Haber, {\em {The CP conserving two Higgs doublet model:
  The Approach to the decoupling limit}}.
  \href{http://dx.doi.org/10.1103/PhysRevD.67.075019}{Phys. Rev. D {\bf 67}
  (2003)  075019}, \href{http://arxiv.org/abs/hep-ph/0207010}{{\tt
  arXiv:hep-ph/0207010}}.

\bibitem{Lee:2012jn}
J.~S. Lee and A.~Pilaftsis, {\em {Radiative Corrections to Scalar Masses and
  Mixing in a Scale Invariant Two Higgs Doublet Model}}.
  \href{http://dx.doi.org/10.1103/PhysRevD.86.035004}{Phys. Rev. D {\bf 86}
  (2012)  035004}, \href{http://arxiv.org/abs/1201.4891}{{\tt arXiv:1201.4891
  [hep-ph]}}.

\bibitem{Coleman:1973jx}
S.~R. Coleman and E.~J. Weinberg, {\em {Radiative Corrections as the Origin of
  Spontaneous Symmetry Breaking}}.
  \href{http://dx.doi.org/10.1103/PhysRevD.7.1888}{Phys. Rev. D {\bf 7} (1973)
  1888--1910}.

\bibitem{Gildener:1976ih}
E.~Gildener and S.~Weinberg, {\em {Symmetry Breaking and Scalar Bosons}}.
  \href{http://dx.doi.org/10.1103/PhysRevD.13.3333}{Phys. Rev. D {\bf 13}
  (1976)  3333}.

\bibitem{Lane:2018ycs}
K.~Lane and W.~Shepherd, {\em {Natural stabilization of the Higgs
  boson{\textquoteright}s mass and alignment}}.
  \href{http://dx.doi.org/10.1103/PhysRevD.99.055015}{Phys. Rev. D {\bf 99}
  (2019) no.~5, 055015}, \href{http://arxiv.org/abs/1808.07927}{{\tt
  arXiv:1808.07927 [hep-ph]}}.

\bibitem{Lane:2019dbc}
K.~Lane and E.~Pilon, {\em {Phenomenology of the new light Higgs bosons in
  Gildener-Weinberg models}}.
  \href{http://dx.doi.org/10.1103/PhysRevD.101.055032}{Phys. Rev. D {\bf 101}
  (2020) no.~5, 055032}, \href{http://arxiv.org/abs/1909.02111}{{\tt
  arXiv:1909.02111 [hep-ph]}}.

\bibitem{Eichten:2021qbm}
E.~J. Eichten and K.~Lane, {\em {Higgs alignment and the top quark}}.
  \href{http://dx.doi.org/10.1103/PhysRevD.103.115022}{Phys. Rev. D {\bf 103}
  (2021) no.~11, 115022}, \href{http://arxiv.org/abs/2102.07242}{{\tt
  arXiv:2102.07242 [hep-ph]}}.

\bibitem{Braathen:2020vwo}
J.~Braathen, S.~Kanemura, and M.~Shimoda, {\em {Two-loop analysis of
  classically scale-invariant models with extended Higgs sectors}}.
  \href{http://dx.doi.org/10.1007/JHEP03(2021)297}{JHEP {\bf 03} (2021)  297},
  \href{http://arxiv.org/abs/2011.07580}{{\tt arXiv:2011.07580 [hep-ph]}}.

\bibitem{Eichten:2022vys}
E.~J. Eichten and K.~Lane, {\em {Gildener-Weinberg two-Higgs-doublet model at
  two loops}}. \href{http://dx.doi.org/10.1103/PhysRevD.107.075038}{Phys. Rev.
  D {\bf 107} (2023) no.~7, 075038},
  \href{http://arxiv.org/abs/2209.06632}{{\tt arXiv:2209.06632 [hep-ph]}}.

\bibitem{Nhi:2025iob}
N.~D.~B. Nhi and E.~Senaha, {\em {Exploring CP violation and vanishing electric
  dipole moment of the electron in a scale-invariant general 2HDM}}.
  \href{http://dx.doi.org/10.1103/hkbf-ddvq}{Phys. Rev. D {\bf 113} (2026)
  no.~1, 015019}, \href{http://arxiv.org/abs/2509.14708}{{\tt arXiv:2509.14708
  [hep-ph]}}.

\bibitem{Baouche:2025jsf}
N.~Baouche and A.~Ahriche, {\em {Phenomenology of the Minimal Scale Invariant
  Two-Higgs-Doublet Model}}. \href{http://arxiv.org/abs/2511.06049}{{\tt
  arXiv:2511.06049 [hep-ph]}}.

\bibitem{Pilaftsis:2024uub}
A.~Pilaftsis, {\em {Dirac algebra formalism for Two Higgs Doublet Models: The
  one-loop effective potential}}.
  \href{http://dx.doi.org/10.1016/j.physletb.2024.139147}{Phys. Lett. B {\bf
  860} (2025)  139147}, \href{http://arxiv.org/abs/2408.04511}{{\tt
  arXiv:2408.04511 [hep-ph]}}.

\bibitem{Degrassi:2023eii}
G.~Degrassi and P.~Slavich, {\em {On the two-loop BSM corrections to
  $h\longrightarrow \gamma \gamma $ in the aligned THDM}}.
  \href{http://dx.doi.org/10.1140/epjc/s10052-023-12097-3}{Eur. Phys. J. C {\bf
  83} (2023) no.~10, 941}, \href{http://arxiv.org/abs/2307.02476}{{\tt
  arXiv:2307.02476 [hep-ph]}}.

\bibitem{Degrassi:2025pqt}
G.~Degrassi, R.~Gr{\"o}ber, and P.~Slavich, {\em {Two-loop BSM contributions to
  Higgs pair production in the aligned THDM}}.
  \href{http://dx.doi.org/10.1007/JHEP01(2026)041}{JHEP {\bf 01} (2026)  041},
  \href{http://arxiv.org/abs/2508.11539}{{\tt arXiv:2508.11539 [hep-ph]}}.

\bibitem{Hahn:2000kx}
T.~Hahn, {\em {Generating Feynman diagrams and amplitudes with FeynArts 3}}.
  \href{http://dx.doi.org/10.1016/S0010-4655(01)00290-9}{Comput. Phys. Commun.
  {\bf 140} (2001)  418--431}, \href{http://arxiv.org/abs/hep-ph/0012260}{{\tt
  arXiv:hep-ph/0012260}}.

\bibitem{Martin:2003it}
S.~P. Martin, {\em {Two loop scalar self energies in a general renormalizable
  theory at leading order in gauge couplings}}.
  \href{http://dx.doi.org/10.1103/PhysRevD.70.016005}{Phys. Rev. D {\bf 70}
  (2004)  016005}, \href{http://arxiv.org/abs/hep-ph/0312092}{{\tt
  arXiv:hep-ph/0312092}}.

\bibitem{Goodsell:2019zfs}
M.~D. Goodsell and S.~Pa{\ss}ehr, {\em {All two-loop scalar self-energies and
  tadpoles in general renormalisable field theories}}.
  \href{http://dx.doi.org/10.1140/epjc/s10052-020-7657-8}{Eur. Phys. J. C {\bf
  80} (2020) no.~5, 417}, \href{http://arxiv.org/abs/1910.02094}{{\tt
  arXiv:1910.02094 [hep-ph]}}.

\bibitem{Marciano:1980pb}
W.~J. Marciano and A.~Sirlin, {\em {Radiative Corrections to Neutrino Induced
  Neutral Current Phenomena in the SU(2)-L x U(1) Theory}}.
  \href{http://dx.doi.org/10.1103/PhysRevD.22.2695}{Phys. Rev. D {\bf 22}
  (1980)  2695}. [Erratum: Phys.Rev.D 31, 213 (1985)].

\bibitem{Sirlin:1985ux}
A.~Sirlin and R.~Zucchini, {\em {Dependence of the Quartic Coupling H(m) on
  M($H$) and the Possible Onset of New Physics in the Higgs Sector of the
  Standard Model}}. \href{http://dx.doi.org/10.1016/0550-3213(86)90096-9}{Nucl.
  Phys. B {\bf 266} (1986)  389--409}.

\bibitem{Degrassi:1992ff}
G.~Degrassi and A.~Sirlin, {\em {Gauge dependence of basic electroweak
  corrections of the standard model}}.
  \href{http://dx.doi.org/10.1016/0550-3213(92)90671-W}{Nucl. Phys. B {\bf 383}
  (1992)  73--92}.

\bibitem{Toussaint:1978zm}
D.~Toussaint, {\em {Renormalization Effects From Superheavy Higgs Particles}}.
  \href{http://dx.doi.org/10.1103/PhysRevD.18.1626}{Phys. Rev. D {\bf 18}
  (1978)  1626}.

\bibitem{Pilaftsis:2011ed}
A.~Pilaftsis, {\em {On the Classification of Accidental Symmetries of the Two
  Higgs Doublet Model Potential}}.
  \href{http://dx.doi.org/10.1016/j.physletb.2011.11.047}{Phys. Lett. B {\bf
  706} (2012)  465--469}, \href{http://arxiv.org/abs/1109.3787}{{\tt
  arXiv:1109.3787 [hep-ph]}}.

\bibitem{Cepeda:2019klc}
M.~Cepeda {\em et al.}, {\em {Report from Working Group 2}: {Higgs Physics at
  the HL-LHC and HE-LHC}}.
  \href{http://dx.doi.org/10.23731/CYRM-2019-007.221}{CERN Yellow Rep. Monogr.
  {\bf 7} (2019)  221--584}, \href{http://arxiv.org/abs/1902.00134}{{\tt
  arXiv:1902.00134 [hep-ph]}}.

\bibitem{Hahn:1998yk}
T.~Hahn and M.~Perez-Victoria, {\em {Automatized one loop calculations in
  four-dimensions and D-dimensions}}.
  \href{http://dx.doi.org/10.1016/S0010-4655(98)00173-8}{Comput. Phys. Commun.
  {\bf 118} (1999)  153--165}, \href{http://arxiv.org/abs/hep-ph/9807565}{{\tt
  arXiv:hep-ph/9807565}}.

\end{thebibliography}\endgroup

\end{document}